\documentclass[fleqn,10pt]{wlscirep}
\usepackage[utf8]{inputenc}
\usepackage[T1]{fontenc}
\usepackage[numbers,sort]{natbib}
\usepackage{bbm,psfrag}

\newcommand{\R}{\mathbb R}
\newcommand{\id}{\mathbbm{1}}
\renewcommand{\P}{\mathbf P}

\renewcommand{\d}{\mathrm d}
\renewcommand{\Re}{\operatorname{Re}}
\newcommand{\GG}{\operatorname{\Gamma G}}
\newcommand{\GGM}{\operatorname{\Gamma GM}}

\title{Modelling the age distribution of longevity leaders}

\author[1]{Csaba Kiss}
\author[2,3,*]{L\'aszl\'o N\'emeth}
\author[1,4]{B\'alint Vet\H{o}}

\affil[1]{Department of Stochastics, Institute of Mathematics,
Budapest University of Technology and Economics, M\H uegyetem rkp.\ 3., H-1111 Budapest, Hungary.}
\affil[2]{Weierstrass Institute for Applied Analysis and Stochastics, Mohrenstraße 39, D-10117 Berlin, Germany.}
\affil[3]{Max Planck Institute for Demographic Research, Konrad-Zuse-Str. 1., D-18057 Rostock, Germany.}
\affil[4]{ELKH--BME Stochastics Research Group, M\H uegyetem rkp.\ 3., H-1111 Budapest, Hungary.}

\affil[*]{nemeth@wias-berlin.de}

\keywords{human longevity, world's oldest person, stochastic model, gamma--Gompertz lifespan distribution}

\begin{abstract}
Human longevity leaders with remarkably long lifespan play a crucial role in the advancement of longevity research.
In this paper, we propose a stochastic model to describe the evolution of the age of the
oldest person in the world by a Markov process, in which we assume that the births of the individuals follow a Poisson process with increasing intensity, lifespans of individuals are independent and can be characterized by a gamma--Gompertz distribution with time-dependent parameters.
We utilize a dataset of the world’s oldest person title holders since 1955, and we compute the maximum likelihood estimate for the parameters iteratively by numerical integration.
Based on our preliminary estimates, the model provides a good fit to the data and shows
that the age of the oldest person alive increases over time in the future.
The estimated parameters enable us to describe the distribution of the age of the record holder process at a future time point.
\end{abstract}

\begin{document}

\maketitle

\flushbottom

\thispagestyle{empty}

\section*{Introduction}

Exceptionally long human lifespans are one of the cornerstones of demography and mortality research.
Studying the group of record holders may reveal not only the underlying mortality mechanism of a population but also potentially shed some light on the future developments of human longevity.
As life expectancy increases \citep{OepVau02} with deaths shifting to older ages, the distribution of deaths at the oldest-old ages \citep{Can10, Vauetal21} gains more interest of demographers, actuaries and decision makers of numerous disciplines.

The pattern of adult human mortality has already been described \citep{Gom25} but there is still a debate about the exact distribution of deaths at adult ages. More details on the necessity of a correctly specified model for the underlying mortality process and its impact on further research are discussed in \citep{NemMis18, Misetal16}. Numerous publications review the ongoing discussion on the existence of a mortality plateau \citep{Maietal21,Danetal23,VijLeb17,Alvetal21,Beletal22,RooZho17}, and the levelling-off of adult human death rates at the oldest ages is supported by the findings in \citep{Baretal18, Modetal17,WilRob03} while others cast some doubt on this observation \citep{GAvGav11,New18,Cam22}. If there is a mortality plateau then the distribution of deaths at the oldest-old ages must be gamma-Gompertz and human lifespan can increase further without any maximum \citep{MisVau15}. Further studies discuss the existence of a limit to human lifespan with more focus on the extreme value distribution aspect of the deaths at the oldest-old ages for various populations and different lifetime distributions \citep{Gbaetal17,HanSib16,Einetal19,LiLiu20,Mil20}. These models can be helpful in determining the plausibility of longevity leaders as well.

We contribute to this discussion by proposing a stochastic model to describe the evolution of the age of the world's oldest person. Based on our estimates the model provides a good fit to the titleholder data since 1955, collected by the Gerontology Research Group \citep{tableurl}.
With the model results, it is possible to predict the age of the oldest person in the world in the future.
When should we expect to see the next Jeanne Calment, the supercentenarian with the longest human lifespan ever documented?
Will her record ever be surpassed?
Our results provide a prediction for the age distribution of the record holder in the coming decades to answer these questions.

\section*{Results}

Our model describes the evolution of the age of the oldest living person under the following assumptions.
We assume that the births of individuals follow a Poisson process with time-dependent intensity \citep{Bri86}.
The lifespans of individuals in the population are independent and their distribution may depend on the date of birth.
Then the age of the record holder in the population evolves in time as a Markov process with explicit transition probabilities.
As the first main result of this paper, we explicitly compute the distribution of the age of the record holder for any given birth rate parameter and lifespan distribution.
The detailed mathematical description and the properties of the general model are described in Section~\ref{s:mathmodel}.

\begin{figure}
\centering
\includegraphics[width=250pt]{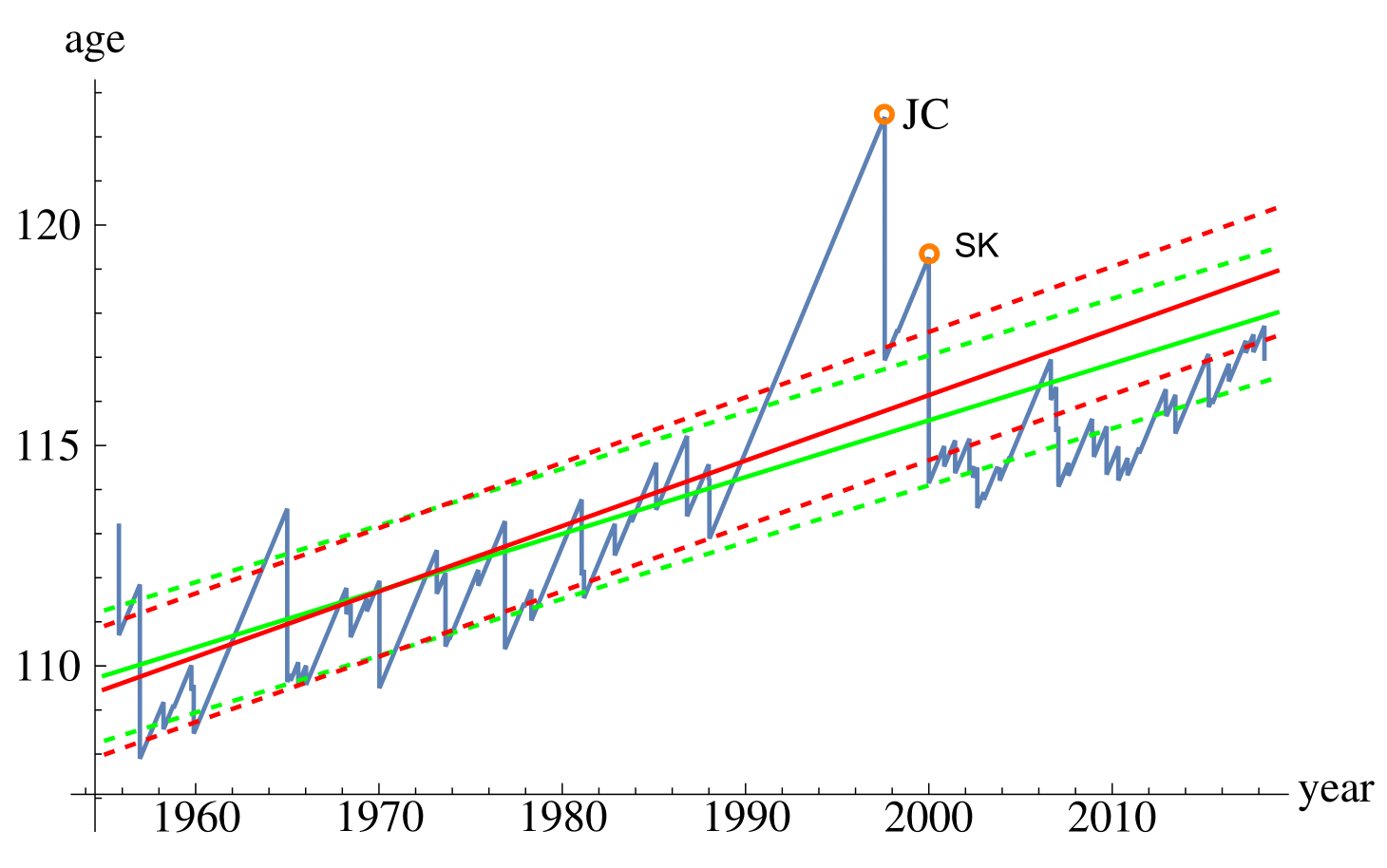}
\caption{Blue curve: age of the oldest person in the world since 1955 with two outliers indicated (Jeanne Calment: JC and Sarah Knauss: SK).
Green curves: the model-estimated mean age of the oldest person (solid line) and the standard deviation (distance from the dashed lines) with the estimation based on the full dataset.
Red curves: mean age and standard deviation of the oldest person estimated using the data between 1955 and 1988.}
\label{fig:backtest}
\end{figure}

We apply our general result to the case which approximates the human birth rate over the world and the human lifespan.
We specify the intensity function of the Poisson process of births to have an exponential growth in time.
The underlying force of mortality is chosen so that it follows an extension of the Gompertz mortality model \citep{Gom25}
and the lifespan distribution of individuals is given by the gamma--Gompertz ($\GG$) distribution with time-dependent parameters.
This distribution adequately captures the slowing down of senescence mortality at the oldest old ages.
Given the growth parameters of the birth rate, we fit the model parameters to the statistics of the oldest person titleholder data using maximum likelihood method.
The optimal parameters of the model fit well to the data.
It shows in particular that the age of the oldest person alive increases over time, and it will most likely increase further in the future.
We compute the expected value and a confidence interval for the age of the world's oldest person using the fitted model parameters for each year between 1955 and 2019 shown by the green curves on Figure~\ref{fig:backtest}.
The detailed discussion of the model specification, likelihood calculations as well as the parameter fitting are given in Section~\ref{s:fitting}.
Section~\ref{s:methods} contains calculations related to the gamma--Gompertz--Makeham generalization of the gamma--Gompertz distribution.

\begin{figure}
\centering
\includegraphics[width=250pt]{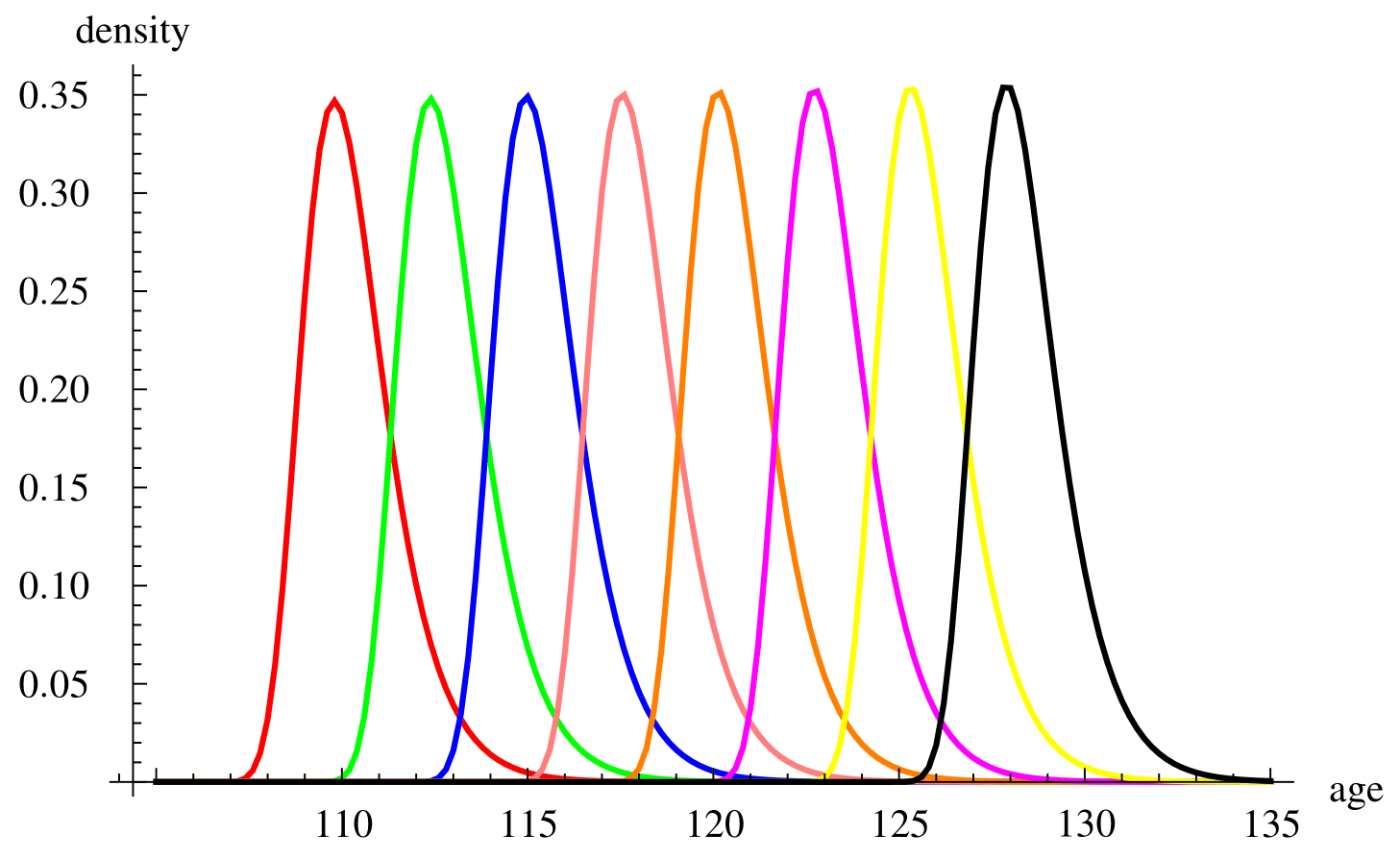}
\caption{The probability densities of the age of the oldest person in the world in the following years: 1960 (red), 1980 (green), 2000 (blue), 2020 (pink), 2040 (orange), 2060 (magenta), 2080 (yellow) and 2100 (black).\label{fig:predictions}}
\end{figure}

Our results enable us to predict the age distribution of the world's oldest person at future time points.
We compute the probability density of the age of the world's oldest person in different years not only in the past but also in the future.
These densities are shown in Figure~\ref{fig:predictions}.
When comparing the age distribution of the oldest person in the world in different years to the age of Jeanne Calment at her death, we find that on Jan 1st 2060 we can expect that the age of the world's oldest person will exceed her age with probability around $0.5$.
This also means that with high probability her age record will already be broken by that time.

In Figure~\ref{fig:backtest} two extreme outliers with unexpectedly long lifetimes can be observed. Jeanne Calment died at the age of 122.45 years in 1997, and Sarah Knauss died at the age 119.27 in 2000.
In our model, the probability of observing an age greater than or equal to their actual age at the time of their death is $0.000286$ for Calment and $0.0116$ for Knauss.
See details in Subsection~\ref{ss:agecomp}.

The fact that Calment and Knauss are outliers among the oldest old in the world became even more evident when we performed a backtesting of our model. We estimated the parameters based on the data on the world's oldest person between 1955 and 1988 where the ending date is the time when Calment became the world's oldest person. The model-estimated mean and confidence interval of the world's oldest person using the full data and the partial dataset (before Calment) are shown in Figure~\ref{fig:backtest} by green and red, respectively.
The estimate using the data until 1988 is less reliable after 2000 which is shown by the fact that the observed data is out of the confidence interval in the majority of the time after 2000.
When we compare the two confidence intervals, we can conclude that, based on the data before 1988, Calment and Knauss already had extremely high ages at their deaths.
Adding the remaining data set, the estimated mean age of the world's oldest person becomes lower.
Hence we can conclude that now we can consider the ages of the two outliers between 1988 and 2000 at their death to be more extreme than based on the information available in 1988.

The other important observable is the reign length of a record holder.
The numerical value of the expected reign length with our estimated model parameters is $1.195$ in 1955 and it is $1.188$ in 2019.
The empirical value of the reign length is $1.008$ which is not much less than the model-based estimate.

Our approach to studying the age of the oldest old is completely novel because it takes into account jointly the age and the time of birth of individuals.
Although the age and the reign length of the world's oldest person depend in a complex non-linear way on the total lifespan and the time of birth of supercentenarians, we compute explicitly the probability distribution of the age of the oldest person.
Hence, the performance of our predictions cannot be directly compared to previous results in the literature because in the usual approach, the oldest person in each cohort is considered separately, and it is not relevant whether this person was ever the oldest in the population, see e.g.\ the extreme value method in~\cite{Gbaetal17}.
Our model contributes to the mathematical understanding of the evolution of the oldest individual, which is the extra benefit compared to a prediction using the trend in the data, e.g.\ in linear regression.
In this way, we not only observe but also prove mathematically that the dynamics of the birth process and that of the lifespan distribution which we consider in this paper necessarily imply the increase of the expected age of the world's oldest person.

\section{Mathematical model for the age of the oldest person}
\label{s:mathmodel}

In this section, we provide the mathematical definition of a general model for the age of the world's oldest person, where the births of individuals follow a Poisson process and their lifespans are independent.
Under the assumptions of the general model, the age of the record holder in the population evolves in time as a Markov process with explicit transition probabilities.
In Subsections~\ref{ss:2Drep}--\ref{ss:exactdistr}, we describe the exact distribution of the age of the record holder in this generality
for any given birth rate parameter and lifespan distribution using the two-dimensional representation of the age process of the oldest person.
In the time-homogeneous case with constant birth rate and identical lifespan distributions the reign length distribution of a record holder is computed in Subsections~\ref{ss:homog}--\ref{ss:reignlength}.
We explain the role of the entry age parameter in Subsection~\ref{ss:entryage}.

\subsection{Model description and two-dimensional representation}
\label{ss:2Drep}

The model is formally defined as follows.
Let $\lambda(t)$ be the birth rate parameter which depends on time and let $F_t$ and $f_t$ be a family of cumulative distribution functions and density functions corresponding to non-negative random variables which are also time-dependent.
We assume that individuals are born according to a Poisson point process at rate $\lambda(t)$ and that the lifespan of an individual born at time $t$ is given by $F_t$ so that lifespans are independent for different individuals.

Let $Y_t$ denote the age of the oldest person in the population at time $t$.
The process $(Y_t:t\in\R)$ is Markovian.
The Markov property holds because at any time $t$ the history of the process $(Y_s:s\le t)$ provides information about the lifetime of individuals born before the current record holder
while any transition of $(Y_s:s\ge t)$ depends only on the lifetime of the current record holder and of those born after them.

The evolution of the Markov process $Y_t$ is the following.
It has a deterministic linear growth with slope $1$ due to the ageing of the current record holder.
This happens until the death of the record holder.
Additionally, given that $Y_{t-}=\lim_{s\uparrow t}Y_s=y$ for some $t$ with $y>0$,
the process has a downward jump at time $t$ at rate $f_t(y)/(1-F_t(y))$ which is the hazard rate of the distribution $F_t$ at $y$.
This corresponds to the possibility that the record holder dies at time $t$ which happens at rate $f_t(y)/(1-F_t(y))$.
The conditional distribution of the jump is given by
\begin{equation}\label{Yttransition}
\P(Y_t<x\,|\,Y_{t-}=y,Y_t<y)
=\exp\left(-\int_x^y\lambda(t-u)(1-F_{t-u}(u))\,\d u\right)
\end{equation}
for all $x>0$.
The jump distribution in \eqref{Yttransition} has an absolutely continuous part supported on $[0,y]$ with density
\begin{equation}\label{defj}
j_{y,t}(x)=\exp\left(-\int_x^y\lambda(t-u)(1-F_{t-u}(u))\,\d u\right)
\lambda(t-x)\left(1-F_{t-x}(x)\right)
\end{equation}
and a point mass at $0$ with probability
\begin{equation}\label{defa}
a_{y,t}=\exp\left(-\int_0^y\lambda(t-u)(1-F_{t-u}(u))\,\d u\right).
\end{equation}
As we shall see in the relevant parameter regime the probability $a_{y,t}$ of the point mass at $0$ is negligible.
The transition formula in \eqref{Yttransition} can be proven using the description below.

We introduce a two-dimensional representation of the process $Y_t$ as follows.
Let $\Lambda=\{(t_i,x_i):i\in I\}$ be a marked Poisson process in $\R\times\R_+$
where $\{t_i:i\in I\}$ forms a Poisson point process on $\R$ with intensity $\lambda(t)$
and $x_i\ge0$ is sampled independently for each $i\in I$ according to the distribution $F_{t_i}$.
The point $(t_i,x_i)$ represents an individual born at time $t_i$ with lifespan $x_i$ for all $i\in I$, that is, the individual $i$ is alive in the time interval $[t_i,t_i+x_i)$ and their age at time $t$ is $t-t_i$ if $t\in[t_i,t_i+x_i)$.
Hence the marked Poisson process $\Lambda$ contains all relevant information about the age statistics of the population at any time.
In particular the age of the oldest person $Y_t$ can be expressed in terms of $\Lambda$ as
\begin{equation}\label{YtfromLambda}
Y_t=\max\left\{(t-t_i)\id_{\{t\in[t_i,t_i+x_i)\}}:i\in I\right\}.
\end{equation}
where the indicator $\id_{\{t\in[t_i,t_i+x_i)\}}$ is $1$ exactly if the $i$th person is alive at time $t$.

\begin{figure}
\centering
\includegraphics[width=250pt]{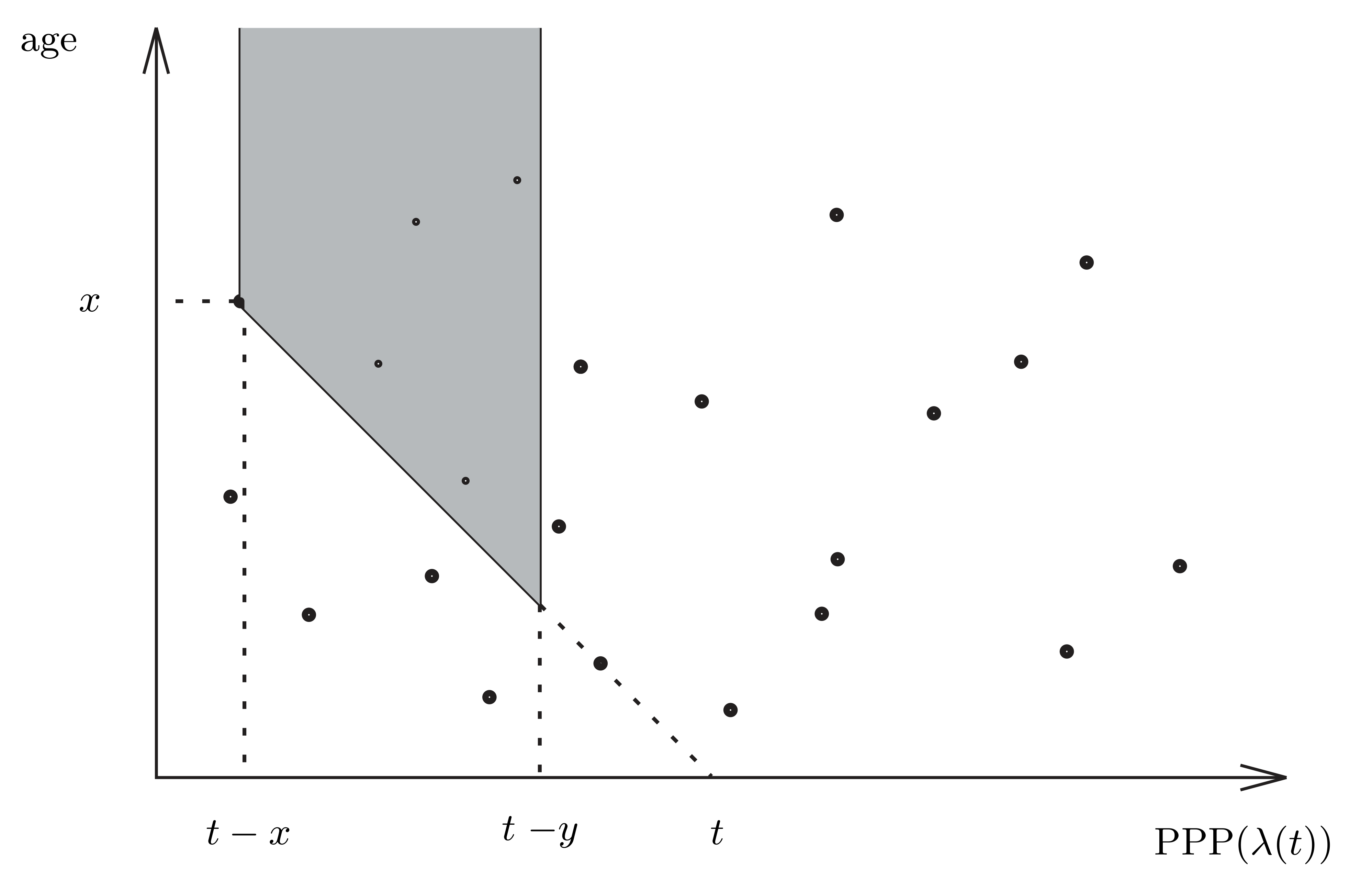}
\caption{Two-dimensional representation of the event $\{Y_t<x\}$ conditionally given $\{Y_{t-}=y,Y_t<y\}$:
the marked Poisson process does not have any point in the grey region, see \eqref{Yttransition}.\label{abra11}}
\end{figure}

The transition distribution formula \eqref{Yttransition} can be seen using the two-dimensional representation as follows.
Given that the current record holder dies at time $t$ at age $y$ the event $\{Y_t<x\}$ means that nobody with age between $x$ and $y$ can be alive at time $t$.
This event can be equivalently characterized in terms of the Poisson process of birth at rate $\lambda(\cdot)$ thinned by the probability that the person is still alive at time $t$.
Indeed the event $\{Y_t<x\}$ can be expressed as a Poisson process of intensity at time $t-u$ given by $\lambda(t-u)(1-F_{t-u}(u))$ for $u\in[x,y]$ not having any point in the time interval $[t-y,t-x]$.
This probability appears exactly on the right-hand side of \eqref{Yttransition}.
In other words, for any $u\in[x,y]$ people are born at time $t-u$ at rate $\lambda(t-u)$.
On the other hand, the probability for a person born at time $t-u$ to be alive at time $t$ (that is, at age $u$) is $1-F_{t-u}(u)$.
See Figure~\ref{abra11} for illustration.

\subsection{Exact distribution of the oldest person's age process}
\label{ss:exactdistr}

We assume that all birth events are already sampled on $(-\infty,t]$ together with the corresponding lifespans.
Then the distribution of $Y_t$ can be computed explicitly for all $t\in\R$ using the two-dimensional representation.
For all $t\in\R$ the density
\begin{equation}\label{statdensity}
h_t(x)=\exp\left(-\int_x^{\infty}\lambda(t-u)(1-F_{t-u}(u))\,\d u\right)
\lambda(t-x)(1-F_{t-x}(x))
\end{equation}
for all $x>0$ and the point mass at $0$
\begin{equation}\label{statmass}
m_t=\exp\left(-\int_0^{\infty}\lambda(t-u)(1-F_{t-u}(u))\,\d u\right)
\end{equation}
characterize the distribution of $Y_t$ which can be seen as follows.
We mention that the point mass at $0$ is negligible in the application.

\begin{figure}
\centering
\includegraphics[width=250pt]{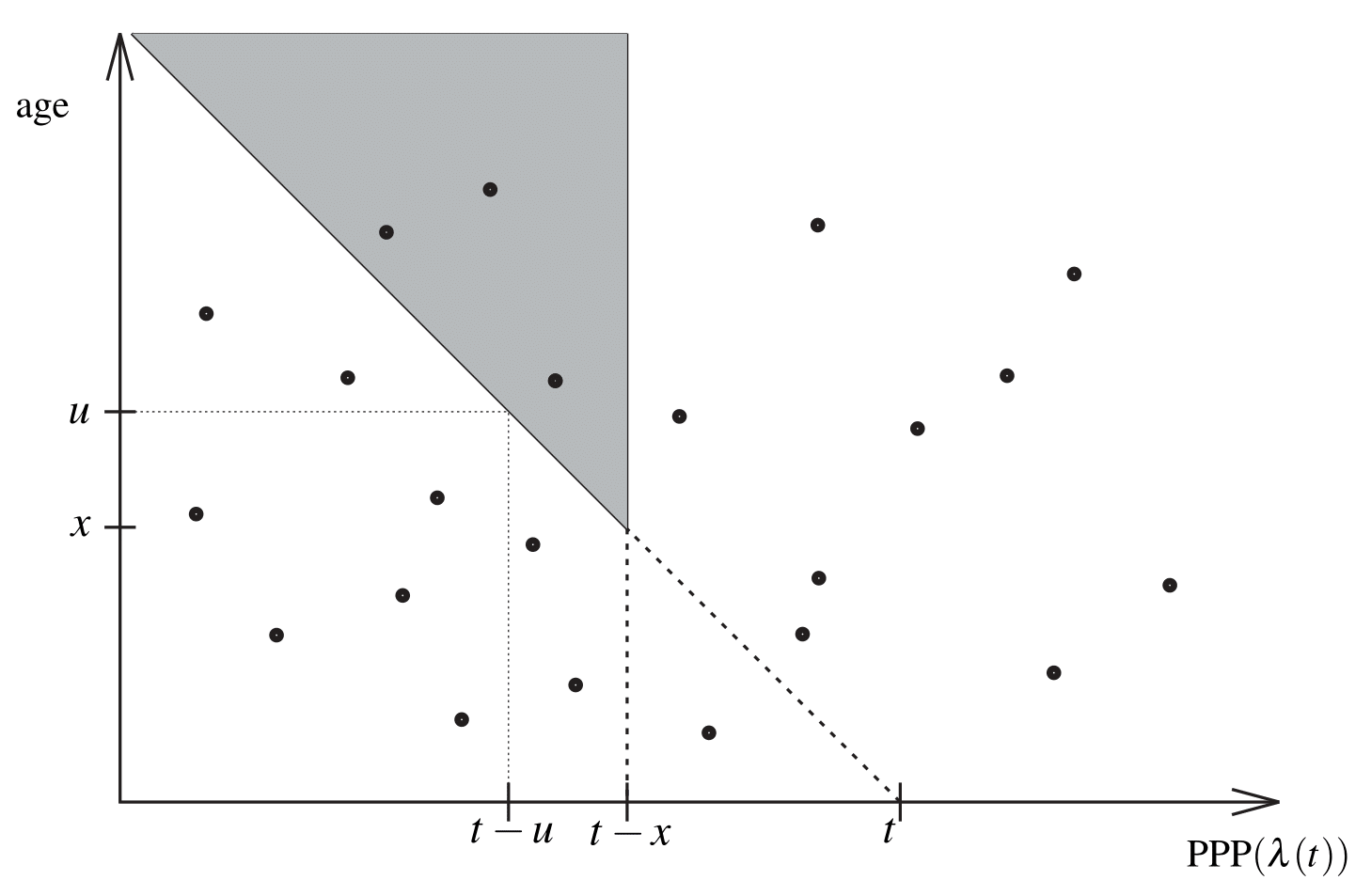}
\caption{Two-dimensional representation of the event $\{Y_t<x\}$: the marked Poisson process does not have any point in the grey region, see \eqref{Ytdistr}.\label{abra1}}
\end{figure}

Similarly to the proof of the transition formula in \eqref{Yttransition} the event $\{Y_t<x\}$ for any $x>0$ is the same as the event that nobody with age at least $x$ is alive at time $t$.
We express this event in terms of the Poisson process of birth at rate $\lambda(\cdot)$ thinned by the probability that the person is still alive at time $t$.
The event $\{Y_t<x\}$ means that a Poisson process of intensity at time $t-u$ given by $\lambda(t-u)(1-F_{t-u}(u))$ for $u\ge x$ does not have any point in $(-\infty,t-x]$ yielding
\begin{equation}\label{Ytdistr}
\P(Y_t<x)=\exp\left(-\int_x^{\infty}\lambda(t-u)(1-F_{t-u}(u))\,\d u\right).
\end{equation}
In other words for any $u\ge x$ individuals are born at time $t-u$ at rate $\lambda(t-u)$.
A person born at time $t-u$ is alive at time $t$ at age $u$ with probability $1-F_{t-u}(u)$.
\eqref{statdensity}--\eqref{statmass} follow by differentiation in \eqref{Ytdistr} and by taking the $x\to0$ limit.
See Figure~\ref{abra1} for illustration.

\subsection{Homogeneous model}
\label{ss:homog}

The exact computation of the reign length distribution (see Subsection~\ref{ss:reignlength}) can only be performed in a special case of our general model described in Subsection~\ref{ss:2Drep}.
We introduce this special case as the homogeneous model where individuals are born at the times of a Poisson process of constant rate $\lambda=\lambda(t)>0$ for all $t$.
The lifespan of individuals are independent and identically distributed with a fixed density $f=f_t$ and cumulative distribution function $F=F_t$ for all $t$ which does not depend on time.

In the homogeneous model the jump distribution given in \eqref{defj}--\eqref{defa} simplifies to
\begin{equation}\label{homjumpdistr}\begin{aligned}
j_y(x)&=j_{y,t}(x)=\exp\left(-\lambda\int_x^y(1-F(u))\,\d u\right)\lambda(1-F(x)),\\
a_y&=a_{y,t}=\exp\left(-\lambda\int_0^y(1-F(u))\,\d u\right).
\end{aligned}\end{equation}
The distribution of $Y_t$ does not depend on time in this case hence it is a stationary distribution as well.
The formulas for the density of $Y_t$ and point mass at $0$ reduce in the homogeneous case to
\begin{equation}\label{Ytdistrhom}\begin{aligned}
h(x)&=\exp\left(-\lambda\int_x^{\infty}(1-F(u))\,\d u\right)\lambda(1-F(x)),\\
m&=\exp\left(-\lambda\int_0^{\infty}(1-F(u))\,\d u\right)
\end{aligned}\end{equation}
where the integral $\int_0^\infty(1-F(u))\,\d u$ is equal to the expected lifespan.

The equilibrium condition for the homogeneous density $h$ can be written as
\begin{equation}\label{hdiffinteq}
h'(x)+h(x)\frac{f(x)}{1-F(x)}-\int_x^\infty h(y)\frac{f(y)}{1-F(y)}j_y(x)\,\d y=0.
\end{equation}
After differentiation and using the fact that
\begin{equation}
\frac{\d}{\d x}j_y(x)=j_y(x)\left(\lambda(1-F(x))-\frac{f(x)}{1-F(x)}\right)
\end{equation}
one can derive from \eqref{hdiffinteq} the second order differential equation
\begin{equation}
h''(x)+\left(\frac{2f(x)}{1-F(x)}-\lambda(1-F(x))\right)h'(x)
+\left(\frac{2f(x)^2}{(1-F(x))^2}+\frac{f'(x)}{1-F(x)}\right)h(x)=0.
\end{equation}
The point mass $m$ at $0$ satisfies
\begin{equation}\
m\lambda=\int_0^\infty h(y)\frac{f(y)}{1-F(y)}j_y(0)\,\d y.
\end{equation}

\subsection{The peaks process}
\label{ss:peakproc}

In the homogeneous model the sequence of peaks in $Y_t$ forms a discrete time Markov chain.
By peak we mean a local maximum of $Y_t$ with value being equal to the lifespan of the last record holder.
Each time the oldest person dies the process $Y_t$ has a peak with a downward jump following it.
Let $Z_n$ denote the age of record holders at which they die which are the values of the peaks of the process $Y_t$.
The sequence $Z_n$ forms a discrete time Markov chain.
The Markov property follows by the fact that ages at death of previous record holders only give information on people born before the current record holder but transitions depend on the lifespan of the current record holder and that of people born after them.

The stationary density of $Z_n$ is given by
\begin{equation}\label{zstatdensity}
z(x)=\frac{f(x)\exp\left(-\lambda\int_x^\infty(1-F(u))\,\d u\right)}{\int_0^\infty f(y)\exp\left(-\lambda\int_y^\infty(1-F(u))\,\d u\right)\d y}.
\end{equation}
The formula can be seen as follows.
To have a record holder who dies at age $x$ there has to be a person who has lifespan $x$ which gives the factor $f(x)$ in the numerator on the right-hand side of \eqref{zstatdensity}.
The exponential factor is by the two-dimensional representation equal to the probability that no people born before the record holder who just died can be alive at the time the record holder dies.
The denominator on the right-hand side of \eqref{zstatdensity} makes $z(x)$ a probability density function.

The density of $Z_n$ can also be characterized by the following description.
It satisfies the integral equation
\begin{equation}\label{zstatcond}
z(x)=\int_0^\infty z(w)\int_0^{\min(x,y)}j_w(y)\frac{f(x)}{1-F(y)}\,\d y\,\d w+\int_0^\infty z(w)a_wf(x)\,\d w
\end{equation}
which comes from the possible transitions of the peak process as follows.
If the previous record holder had a total lifetime $w\in[0,\infty)$ then at the death the process $Y_t$ jumps down to some value $y$ at rate $j_w(y)$ or to $0$ with probability $a_w$.
The density of the age at which a person dies who becomes a record holder at age $y$ is $f(x)/(1-F(y))$.
From the integral equation in \eqref{zstatcond} one can derive the second order differential equation for the function $g(x)=z(s)/f(x)$ given by
\begin{equation}
g''(x)-\lambda(1-F(x))g'(x)+\lambda f(x)g(x)=0
\end{equation}
which is satisfied by $g(x)=c\exp\left(-\lambda\int_x^\infty(1-F(u))\,\d u\right)$ in accordance with \eqref{zstatdensity}.

\subsection{Reign length distribution}
\label{ss:reignlength}

In the homogeneous model, let $W_n$ denote the reign length of the $n$th record holder, that is, the time length for which this person is the oldest person of the population.
The density of the random reign length is given by
\begin{equation}\label{rreigndensity}
r(w)=\frac{\int_0^\infty h(y)f(y+w)\,\d y+m\int_0^wf(z)\lambda e^{-\lambda(w-z)}f(z)\,\d z}{\int_0^\infty h(y)(1-F(y))\,\d y+m}.
\end{equation}
The density formula in \eqref{rreigndensity} can be derived based on the stationary density of the peaks process given by \eqref{zstatdensity} as follows.

It holds for the density of the reign length that
\begin{equation}\label{ridentity}
r(w)=\int_0^\infty z(x)\left(\int_0^xj_x(y)\frac{f(y+w)}{1-F(y)}\,\d y
+a_x\int_0^wf(z)\lambda e^{-\lambda(w-z)}\,\d z\right)\d x
\end{equation}
based on the decomposition with respect to the previous value of the peaks process $Z_n$.
The integral $\int_0^wf(z)\lambda e^{-\lambda(w-z)}\,\d z$ is the density of the convolution of the density $f$ with an independent exponential distribution of parameter $\lambda$.
On the right-hand side of \eqref{ridentity}, one can use the definitions of the density $z$ given by \eqref{zstatdensity} and the homogeneous jump distribution $j_x$ and $a_x$ given by \eqref{homjumpdistr}.
Then in the numerator after the exchange of the order of integrations in the first term and by using the formula for the stationary distribution given by \eqref{Ytdistrhom} one gets the numerator of \eqref{rreigndensity}.
In the denominator one can use the equality
\begin{equation}
\int_0^\infty f(y)\exp\left(-\lambda\int_y^\infty(1-F(u))\,\d u\right)\d y
=\int_0^\infty h(y)(1-F(y))\,\d y+m.
\end{equation}
which follows by integration by parts.

Note also that the density is not equal to the remaining reign length of $Y_t$ under the stationary distribution because it would involve the integral $\int_0^\infty h(y)f(y+w)/(1-F(y))\,\d y$ in place of the first term in the numerator on the right-hand side of \eqref{rreigndensity}.

\subsection{Entry age parameter}
\label{ss:entryage}

Next we introduce another parameter which we call the entry age and we denote it by $E$.
As opposed to our original model we consider individuals as being born at age $E$ at a modified birth rate and with a modified lifespan distribution.
As a result we obtain a model to the age of the world's oldest person all values of the entry age parameter $E\ge0$ and we can fit the model parameters with different values of the entry age.

For any value $E\ge0$ of the entry age we denote by $\lambda_E(t)$ the rate at which people reach the age $E$ at time $t$, that is,
\begin{equation}\label{lambdaE}
\lambda_E(t)=\lambda(t-E)(1-F_{t-E}(E))
\end{equation}
because the new birth process of rate $\lambda_E(t)$ is obtained by an inhomogeneous thinning of the original Poisson process of the birth events.
The lifespan distribution of those born at time $t$ with age $E$ becomes the remaining lifetime distribution at age $E$.
The modified cumulative distribution function and density are given by
\begin{equation}\label{FtE}
F_t^E(x)=\frac{F_{t-E}(x+E)-F_{t-E}(E)}{1-F_{t-E}(E)},\qquad
f_t^E(x)=\frac{f_{t-E}(x+E)}{1-F_{t-E}(E)}.
\end{equation}

\section{Model specification and parameter fitting}
\label{s:fitting}

In Subsection~\ref{ss:specification} we specify the general model introduced and discussed in Section~\ref{s:mathmodel}, that is,
we assume that the birth rate parameter increases exponentially in time and that the lifespan distribution is given by the gamma--Gompertz--Makeham distribution with time-dependent parameters.
We provide the details of the computation of the likelihood as a function of the model parameters in Subsection~\ref{ss:likelihood}.
We show the way to maximize the likelihood and how the optimal parameters can be found using the Nelder--Mead method in Subsection~\ref{ss:maximization}.
With these values of the parameters, the age of the world's oldest person and the reign length of the record holder can be computed as described in Subsection~\ref{ss:agecomp}.

\subsection{Model specification: birth rate parameter and lifespan distribution}
\label{ss:specification}

For the rest of the paper we specifiy our general model described in Section~\ref{s:mathmodel}
to the following choice of the birth rate parameter and of the lifespan distribution.
We choose the value of the entry age to be $E=0,30,60$ and we fit the model parameters for all three values of $E$ separately.

First we specify the intensity function of the Poisson process of births with an exponential growth in time.
For any of the three values of the entry age $E$ we assume that the birth rate at age $E$ is given by
\begin{equation}\label{lambdat}
\lambda_E(t)=C_Ee^{\kappa_Et}
\end{equation}
where the numerical values of the parameters $C_E$ and $\kappa_E$ are obtained
by linear regression of the logarithmic data of newborns, people at age $30$ and $60$ published by the United Nations since 1950.
We extrapolate the linear regression backwards in time and we use the numerical values shown in Table~\ref{table:birthrate}.

We assume that the underlying force of mortality is chosen so that the lifespan distribution of individuals follows the gamma--Gompertz ($\GG$) distribution with cumulative distribution function and density
\begin{equation}\label{defGG}\begin{aligned}
F^{\GG}_{a,b,\gamma}(x)&=1-\left(1+\frac{a\gamma}b(e^{bx}-1)\right)^{-1/\gamma},\\
f^{\GG}_{a,b,\gamma}(x)&=ae^{bx}\left(1+\frac{a\gamma}b(e^{bx}-1)\right)^{-1-\frac1\gamma}
\end{aligned}\end{equation}
for $x\ge0$ where $a,b,\gamma$ are positive parameters.
We mention that the gamma--Gompertz--Makeham ($\GGM$) distribution differs from the $\GG$ distribution by the presence of a non-negative extrinsic mortality parameter $c$ which appears as an additive term in the force of mortality.
See \eqref{defGGM} for the definition of the $\GGM$ distribution.
In our model, we exclude the extrinsic mortality for the following two reasons.
Since the extrinsic mortality becomes irrelevant at high ages and we aim to model the front-end of the death distribution at the oldest-old ages,
we do not expect to obtain a reliable estimate on the extrinsic mortality using the data about the world's oldest person.
On the other hand, as explained later in Subsection~\ref{ss:maximization}, the likelihood maximization provides unrealistic lifespan distributions even for the $\GG$ model if one tries to optimize in all the parameters at the same time.
In order for the algorithm to result in a distribution close to the actual human lifespan distribution, the number of model parameters had to be decreased.

For our model we suppose that in the $\GG$ lifespan distribution, parameters $b=b_E$, the rate of aging and $\gamma=\gamma_E$, the magnitude of heterogeneity
are constants over time and that they only depend on the value of the entry age parameter $E$.
The parameter $a$, the initial level of mortality at the entry age for individuals born at time $t$, depends on time given by the exponentially decreasing function
\begin{equation}\label{defat}
a_E(t)=K_Ee^{-\alpha_E(t-2000)}
\end{equation}
where the exponent $\alpha_E$ and the constant $K_E$ only depends on the entry age $E$.
The reason for subtracting $2000$ in \eqref{defat} is only technical, the numerical values of the parameters do not become tiny with this definition.

In the model with entry age $E$, we assume that the birth rate $\lambda_E(t)$ is given by \eqref{lambdat} and we fit the gamma--Gompetz distribution with parameters $b_E,\gamma_E$ and $a_E(t)$ given by \eqref{defat} for the modified distribution function $F_t^E(x)$ and density $f_t^E(x)$ in \eqref{FtE}.
This means that we search for the best fitting values of the parameters $\alpha_E,K_E,b_E,\gamma_E$ which results in an approximation of the remaining lifetime distribution at the age $E$.

\subsection{Likelihood calculations}
\label{ss:likelihood}

The aim of the maximum likelihood method is to give an estimate to the parameters $\alpha_E$, $K_E$, $b_E$ and $\gamma_E$ for $E=0,30,60$ by finding those values for which the likelihood of the full sample is the largest.
The sample is obtained from the the historical data on the world's oldest person available in~\cite{tableurl}.
We transform this information into a list of triples of the form $(t_i,y_i,z_i)$ for $i=1,\dots,n$
where $t_i$ is the $i$th time in the sample when the oldest person dies at age $y_i$ and the new record holder has age $z_i$ at time $t_i$.
Then the data has to satisfy the consistency relation $t_i-z_i=t_{i+1}-y_{i+1}$ since the two sides express the date of birth of the same person.

In the model with entry age $E$, the likelihood of the $i$th data point $(t_i,y_i,z_i)$ given the previous data point is equal to
\begin{multline}\label{likely}
\frac{f_{t_i-y_i+E}^E(y_i-E)}{1-F_{t_i-y_i+E}^E(z_{i-1}-E)}j^E_{y_i,t_i}(z_i)\\
=\frac{f_{t_i-y_i+E}^E(y_i-E)}{1-F_{t_i-y_i+E}^E(z_{i-1}-E)}
\exp\left(-\int_{z_i}^{y_i}\lambda_E(t_i-u+E)(1-F_{t_i-u+E}^E(u-E))\,\d u\right)
\lambda_E(t_i-z_i+E)(1-F_{t_i-z_i+E}^E(z_i-E))
\end{multline}
for all $i=2,3,\dots,n$ except for $i=1$ in which case the $1-F_{y_1-t_1+E}^E(z_0-E)$ factor in the denominator is missing.
In \eqref{likely} above we use the transition probabilities of the model with entry age $E$ given by
\begin{equation}\label{defjE}
j^E_{y,t}(x)=\exp\left(-\int_x^y\lambda_E(t-u+E)(1-F_{t-u+E}^E(u-E))\,\d u\right)\lambda_E(t-x+E)(1-F_{t-x+E}^E(x-E))
\end{equation}
as a generalization of \eqref{defj}.
The explanation of the left-hand side of \eqref{likely} is that the person died at time $t_i$ at age $y_i$ had age $E$ at time $t_i-y_i+E$.
The previous data point ensures that this person has already reached age $z_{i-1}$ hence we condition their lifetime distribution on this fact.
The transition probabilities in \eqref{defjE} are obtained similarly to \eqref{defj} with the difference that a person at age $u$ with $u\in[x,y]$ at time $t$ had age $E$ at time $t-u+E$.

Note that when computing the likelihood of the full data by multiplying the right-hand side of \eqref{likely} for different values of $i$ the consistency relation of the data implies that the factor $1-F_{t_i-z_i+E}(z_i-E)$ of the $i$th term cancels with the factor $1-F_{t_{i+1}-y_{i+1}+E}(z_i-E)$ coming from the $(i+1)$st term.
Hence the log-likelihood of the full sample is given by
\begin{equation}\label{loglikelihood}\begin{aligned}
&l(\alpha,K,b,\gamma)\\
&=\sum_{i=1}^n\left(\log f_{t_i-y_i+E}^E(y_i-E)
-\int_{z_i}^{y_i}\lambda_E(t_i-u+E)(1-F_{t_i-u+E}^E(u-E))\,\d u+\log\lambda_E(t_i-z_i+E)\right)
+\log(1-F_{t_n-z_n-E}^E(z_n+E))\\
&=\sum_{i=1}^n\left(\log f^{\GG}_{Ke^{-\alpha(t_i-y_i+E-2000)},b,\gamma}(y_i-E)
-\int_{z_i}^{y_i}Ce^{\kappa(t_i-u+E)}\left(1-F^{\GG}_{Ke^{-\alpha(t_i-u+E-2000)},b,\gamma}(u-E)\right)\d u\right)\\
&\qquad+\log\left(1-F^{\GG}_{Ke^{-\alpha(t_n-z_n+E-2000)},b,\gamma}(z_n-E)\right)
+n\log C+\sum_{i=1}^n\kappa(t_i-z_i+E).
\end{aligned}\end{equation}
where we suppress the dependence of the parameters $\alpha,K,b,\gamma$ on the entry age.
Note that the last two terms do not depend on the parameters $\alpha,K,b,\gamma$ hence we can omit these terms in the maximization of the log-likelihood.

\subsection{Likelihood maximization}
\label{ss:maximization}

We implemented the calculation of the log-likelihood function $l(\alpha,K,b,\gamma)$ given by \eqref{loglikelihood} in Python.
We used numerical integration to obtain the integrals on the right-hand side of \eqref{loglikelihood}.
We mention that the general integral formula in \eqref{GGMint} could not be used because the parameter $a$ of the gamma--Gompertz distribution in the integrand depends on the integration variable on the right-hand side of \eqref{loglikelihood}.

In order to maximize the value of the log-likelihood function $l(\alpha,K,b,\gamma)$ we applied the Nelder--Mead method
\cite{NM65} which is already implemented in Python.
We mention that initially we used the gamma--Gompertz--Makeham distribution as lifespan distribution, see \eqref{defGGM} for the definition, which contains the extra parameter $c$ to be fitted but it turned out that the number of model parameters has to be reduced.
The behaviour of the optimization algorithm in the five parameters $\alpha,K,b,c,\gamma$ using the gamma--Gompretz--Makeham model was very similar the case of four parameters $\alpha,K,c,\gamma$ in the gamma--Gompertz model.
Running the optimization in the full set of parameters ($\alpha,K,b,c,\gamma$ in the gamma--Gompertz--Makeham model or $\alpha,K,b,\gamma$ in the gamma--Gompertz model), it turned out that after a few rounds the parameter $K$ started to decrease dramatically and reached values below $10^{-10}$.
The resulting lifespan distribution seemed very unrealistic with almost no mortality before the age of 100.
This happened for all values of the entry age $E=0,30,60$.

We explain this phenomenon by the fact that historical data about the oldest person in the world only gives information about the behaviour of the lifespan distribution between the ages 107 and 123.
The simple optimization in the four parameters $\alpha,K,b,\gamma$ simultaneously yields an excellent fit for the tail decay of the lifespan distribution with the historical data but the result may be very far from the actual human lifespan.
This would limit the practical relevance of our results.

The mathematical reason for the fact that the four-parameter optimization does not result in a satisfactory approximation to the human lifespan distribution is the following.
In these cases, the optimization procedure diverges to those regimes of the parameter space $\R_+^4$ where the corresponding gamma--Gompertz distribution is degenerate.
One can prevent reaching these unrealistic combinations of parameters by reducing the amount of freedom in the optimization.
Hence we specify some of the parameters a priori and we perform the optimization in the remaining ones so that it provides a good fit to the data on the age of the oldest old as well as a realistic lifespan distribution.

We believe that the most robust of the four parameters of the $\GG$ model is $b$ which is the exponent in the time dependence of the mortality rate.
By setting the rate of aging $b=0.09$ the algorithm gives the optimal triple $\alpha,K,\gamma$ with the best likelihood which is very stable under changing the initial values of these parameters.
The running time is also very short.

The Nelder--Mead algorithm, being a numerical maximization method, heavily relies on the tolerance parameter, which determines the minimal improvement required for the algorithm to continue running. If this parameter is set too high, the algorithm might stop before reaching the optimum. Conversely, if set too low, the algorithm might take excessively long to converge. To address this, we drew inspiration from dynamic learning rate algorithms used in neural network training and developed the following meta-algorithm.

First, we run the Nelder--Mead optimization. Based on the improvement from the starting point, we dynamically adjust the tolerance factor, similar to how learning rates are modified during neural network training. We then run the optimization again, recalibrating the tolerance factor based on the observed improvement, and repeat the process. This iterative adjustment allows us to get closer to the optimum, a hypothesis supported by our practical experience with this meta-algorithm.
Following this meta-algorithm, only a few calls of the Nelder--Mead method is enough to reach the optimum.
The Python codes for the likelihood calculations as well as the Nelder--Mead optimization implemented to this problem are available in \cite{Kiss24}.

\begin{figure}
\centering
\includegraphics[width=250pt]{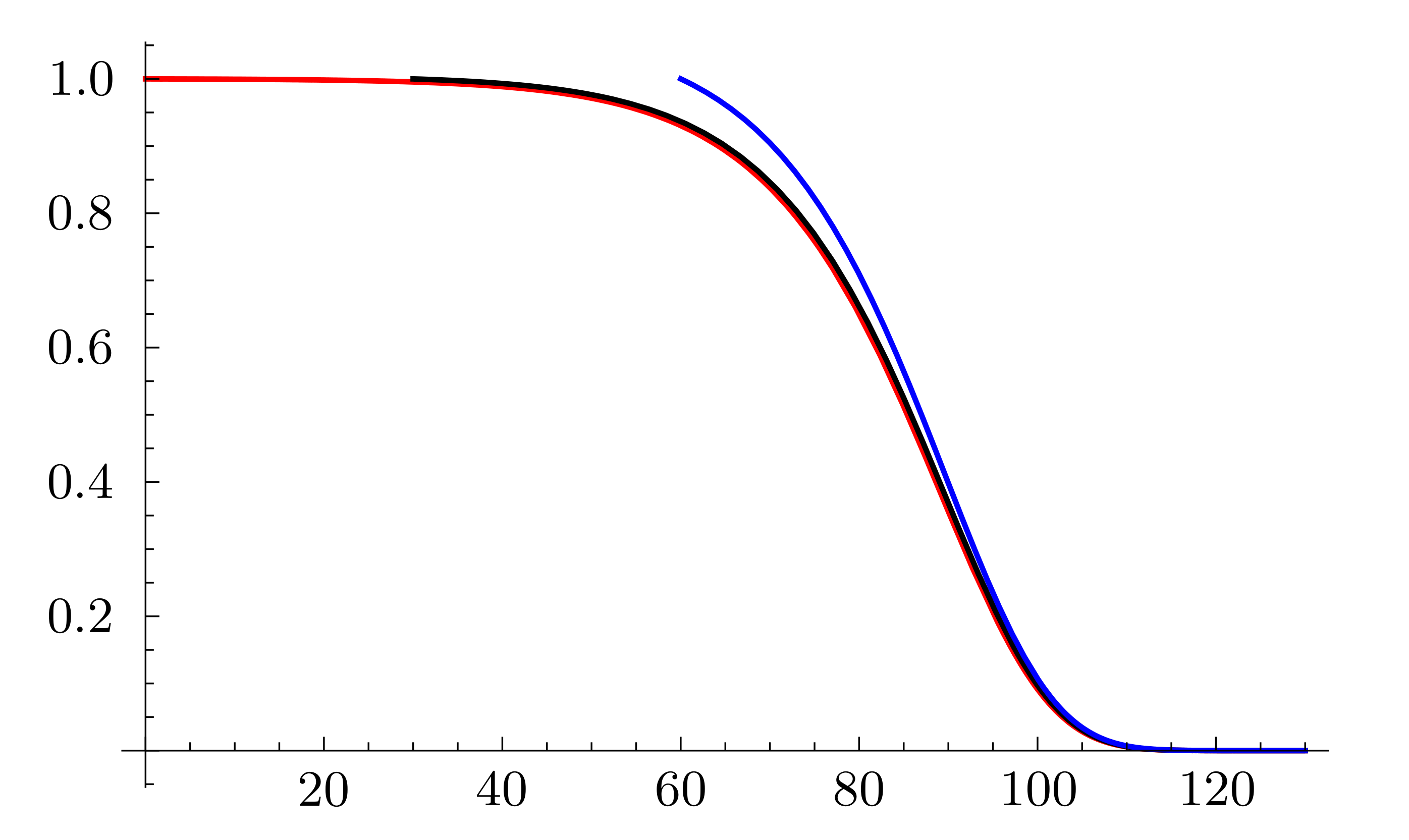}
\caption{Survival probabilities as a function of the age for different values of the entry age: E=0 in red, E=30 in black, E=60 in blue.\label{fig:survival}}
\end{figure}

The numerical values of the resulting parameters for the three choices of the entry age are shown in Table~\ref{table:optparam}.
The survival probability functions with the parameters given in Table~\ref{table:optparam} for individuals born in 2000 corresponding to the entry age $E=0,30,60$ are shown on Figure~\ref{fig:survival} as a function of the age.
We also computed the optimal values of the parameters $\alpha,K,\gamma$ for other values of the rate of aging $b$ as a sensitivity analysis.
The resulting parameter values for the choices $b=0.11$, $b=0.13$ and $b=0.15$ are shown in Table~\ref{table:differentb}.

We mention as an alternative approach that scaling the parameters could enhance the optimization process, but this requires prior knowledge of the range within which the parameters vary. This range could be determined through our iterative application of the Nelder--Mead algorithm.

\subsection{Computation of the oldest person's age and of the reign length}
\label{ss:agecomp}

We observe that the model with the parameters in Table~\ref{table:optparam} fits well to the titleholder data.
We focus on two statistics of the process in order to support this observation about the comparison:
the age of the world's oldest person and the reign length of the record holder.
In the case of both statistics exact formulas are only available for the homogeneous model introduced in Subsection~\ref{ss:homog} where the birth rate is constant as well as the lifespan distribution does not depend on time.
Hence we apply an approximation where the error is negligible compared to the difference from the statistics computed using the data.

In the general model the distribution of the age of the world's oldest person at time $t$ is given by the density $h_t(x)$ in \eqref{statdensity} and by the point mass $m_t$ at $0$ in \eqref{statmass}.
For the numerical computations, we ignore the point mass $m_t$ which is below the round-off error in the numerical results.
The difficulty in computing the mean age of the oldest person at time $t$ is that parameter $a$ of the gamma--Gompertz--Makeham distribution function $F_{t-u}$ in the exponent of \eqref{statdensity} also depends on the integration variable $u$.

In our approximation we fix the value of the parameter $a$ of the $\GG$ distribution in $h_t(x)$ in \eqref{statdensity} to a value which is equal to $a_0(t-d)$ in \eqref{defat} with some delay $d$.
The delay $d$ is chosen so that the mean age of the oldest person computed using $a_0(t-d)$ as parameter $a$ for all times in the $\GG$ distribution function in \eqref{statdensity} is equal to the same value $d$.
For a given $t$, this value of $d$ can be obtained as the fixed point of the contraction map
\begin{equation}\label{iteration}
d\mapsto\int_0^\infty xe^{-\int_x^\infty\lambda(t-u)(1-F_{t-d}(u))\,\d u}\lambda(t-x)(1-F_{t-d}(x))\,\d x
\end{equation}
which provides a reasonable approximation for the mean age of the world's oldest person.
For the comparison with the data and for the prediction, we use the model with entry age $E=0$.
Hence in \eqref{iteration}, the function $\lambda$ is given in \eqref{lambdat} with $E=0$ and $F_{t-d}$ is the $\GG$ distribution function with parameters given by the $E=0$ values in Table~\ref{table:optparam} and with $a=a_0(t-d)$ in \eqref{defat}.
This approximation is reasonable because the distribution of the age of the oldest person is highly concentrated.

The fixed point of the map in \eqref{iteration} as the expected age of the world's oldest person can be found in a few steps of iterations.
We show the result on Figure~\ref{fig:backtest}.
We applied 10 iterations using the fitted model parameters for each year between 1955 and 2019.
By computing the standard deviation of the age of the oldest person as well we obtain the mean and a confidence interval for the age.

The predictions for the age distribution of the world's oldest person in the future shown on Figure~\ref{fig:predictions}.
We obtained them by using the exact age distribution formula in Subsection~\ref{ss:exactdistr} along with the numerical values of the parameters $C,\kappa,\alpha,K,\gamma$ given in Tables~\ref{table:birthrate} and \ref{table:optparam} for entry age $E=0$.

In our model, the distribution function of the age of the world's oldest person is given in \eqref{Ytdistr} where the $\GG$ distribution function can be substituted with the estimated parameter values at any time.
In this way, the probability of observing an age greater than or equal to Calment's or Knauss' actual age at the time of their death can be computed exactly.
The numerical values are $0.000286$ for Calment and it is $0.0116$ for Knauss.

The backtesting mentioned in the Results section is performed as follows.
We estimated the best parameter values with entry age $0$ based on the reduced data on the world's oldest person between 1955 and 1988 where the ending date is the time when Calment became the world's oldest person.
The resulting parameters $\alpha=0.01516,K=0.00002064,\gamma=0.08413$ are numerically not very far from the optimal parameters in Table~\ref{table:optparam} but the difference is more visible on Figure~\ref{fig:backtest}.
The figure shows the model-based mean age and confidence interval for the age of the world's oldest person computed using the full data as well as the data until 1988.

For the reign length of record holders, we again used the expected age at a given time obtained as the fixed point of the iteration in \eqref{iteration}.
The numerical value of the expected reign length obtained from the iteration is $1.195$ in 1955 and it is $1.188$ in 2019.
The empirical value of the reign length is $1.008$ computed from the data by dividing the total length of the time interval between 1955 and 2019 by the number of record holders.

\section{Methods}
\label{s:methods}

In this section, we provide supplementary information related to the main result of this paper.
We perform explicit computations with the gamma--Gompertz--Makeham model
and we express the integral of the survival function in terms of a hypergeometric function.

The cumulative distribution function and the density of the gamma--Gompertz--Makeham ($\GGM$) distribution are given by
\begin{equation}\label{defGGM}\begin{aligned}
F^{\GGM}_{a,b,c,\gamma}(x)&=1-\frac{e^{-cx}}{\left(1+\frac{a\gamma}b(e^{bx}-1)\right)^{1/\gamma}},\\
f^{\GGM}_{a,b,c,\gamma}(x)&=\frac{e^{-cx}}{\left(1+\frac{a\gamma}b(e^{bx}-1)\right)^{1+\frac1\gamma}}\frac{c(b-a\gamma)+a(b+c\gamma)e^{bx}}b
\end{aligned}\end{equation}
for $x\ge0$ where $a,b,c,\gamma$ are positive parameters.
The positivity of parameters implies the finiteness of all moments and, in particular, the convergence of the integral of the survival function
$\int_x^\infty(1-F^{\GGM}_{a,b,c,\gamma}(u))\,\d u$.
In the homogeneous model, the integral of the survival function appears in the density of the distribution of $Y_t$ in \eqref{Ytdistrhom} and in the stationary density of the peaks process in \eqref{zstatdensity}.
We show below that in the gamma--Gompertz--Makeham model the integral of the survival function can be computed explicitly and it is given by
\begin{equation}\label{GGMint}
\int_x^\infty\left(1-F^{\GGM}_{a,b,c,\gamma}(u)\right)\d u
=\left( \frac{b}{a \gamma} \right)^{1/\gamma} \frac{e^{-(c+\frac{b}{\gamma})x}}{\frac{b}{\gamma}+c}
{}_2F_1\left(\frac1\gamma,\frac1\gamma+\frac cb;1+\frac1\gamma+\frac cb;\frac{a \gamma-b}{a \gamma}e^{-bx}\right)
\end{equation}
where $_2F_1(a,b;c;z)$ is the hypergeometric function.
See 15.1.1 in~\cite{AbrSte84} for the definition and properties.

We prove \eqref{GGMint} based on the following integral representation 15.3.1 in~\cite{AbrSte84} of the hypergeometric function
\begin{equation}\label{hypergeomid}
_2F_1(a,b,c,z)=\frac{\Gamma(c)}{\Gamma(b) \Gamma(c-b)}\int_0^1 t^{b-1}(1-t)^{c-b-1}(1-tz)^{-a}\,\d t
\end{equation}
which holds whenever $\Re(c)>\Re(b)>0$.
First we prove an identity for complex parameters $\alpha,\beta,\delta$ which satisfy $\Re(\alpha+\beta)>0$ and we compute
\begin{equation}\label{hypergeomcomp}\begin{aligned}
\int_x^\infty\frac{e^{-\beta u}}{(1+\delta e^u)^\alpha}\,\d u
&=\frac1{\delta^\alpha}\int_x^\infty\frac{e^{-(\alpha+\beta)u}}{(1+e^{-u}/\delta)^\alpha}\,\d u\\
&=\frac{e^{-(\alpha+\beta)x}}{\delta^\alpha}\int_0^1 y^{\alpha+\beta-1}\left(1+e^{-x}y/\delta\right)^{-\alpha}\,\d y\\
&=\frac{e^{-(\alpha+\beta)x}}{(\alpha+\beta)\delta^\alpha}\cdot{}_2F_1\left(\alpha,\alpha+\beta;1+\alpha+\beta;-\frac{e^{-x}}\delta\right)
\end{aligned}\end{equation}
where we applied a change of variables $y=e^{x-u}$ in the second equality above and we applied the hypergeometric identity \eqref{hypergeomid} in the last equality with $a=\alpha$, $b=\alpha+\beta$, $c=1+\alpha+\beta$, $z=-e^{-x}/\delta$ together with the observation that with these values of the parameters the prefactor of the integral on the right-hand side of \eqref{hypergeomid} simplifies to $\alpha+\beta$.
Note that the condition $\Re(c)>\Re(b)>0$ for \eqref{hypergeomid} to hold is satisfied by our assumption $\Re(\alpha+\beta)>0$ which also makes the integrals in \eqref{hypergeomcomp} convergent.

Next we show \eqref{GGMint} using \eqref{hypergeomcomp} as follows.
We write
\begin{equation}\label{hypergeomcomp2}\begin{aligned}
\int_x^\infty\left(1-F^{\GGM}_{a,b,c,\gamma}(u)\right)\,\d u
&=\frac1{\left(1-\frac{a\gamma}b\right)^{1/\gamma}}\int_x^\infty\frac{e^{-cu}}{\left(1+\frac{a\gamma}{b-a\gamma}e^{bu}\right)^{1/\gamma}}\,\d u\\
&=\frac1{\left(1-\frac{a\gamma}b\right)^{1/\gamma}b}\int_{bx}^\infty\frac{e^{-cv/b}}{\left(1+\frac{a\gamma}{b-a\gamma}e^v\right)^{1/\gamma}}\,\d v\\
&=\frac1{\left(1-\frac{a\gamma}b\right)^{1/\gamma}b}\frac{e^{-\left(\frac1\gamma+\frac cb\right)bx}}{\left(\frac1\gamma+\frac cb\right)\left(\frac{a\gamma}{b-a\gamma}\right)^{1/\gamma}}
{}_2F_1\left(\frac1\gamma,\frac1\gamma+\frac cb;1+\frac1\gamma+\frac cb;\frac{a \gamma-b}{a \gamma}e^{-bx}\right)
\end{aligned}\end{equation}
where we applied the change of variables $v=bu$ in the second equality above and we used \eqref{hypergeomcomp} with $\alpha=1/\gamma$, $\beta=c/b$, $\delta=a\gamma/(b-a\gamma)$ and with $x$ replaced by $bx$.
The right-hand side of \eqref{hypergeomcomp2} simplifies to that of \eqref{GGMint}.

\section*{Acknowledgements}

We thank Katalin Kov\'acs for some useful advice at an early stage of the project which led to this collaboration.
The work of Cs.\ Kiss and B.\ Vet\H o was supported by the NKFI (National Research, Development and Innovation Office) grant FK142124.
B.\ Vet\H o is also grateful for the support of the NKFI grant KKP144059 ``Fractal geometry and applications'' and for the Bolyai Research Scholarship of the Hungarian Academy of Sciences.
L. Németh was supported by MaRDI, funded by the Deutsche Forschungsgemeinschaft (DFG), project number 460135501, NFDI 29/1 “MaRDI -- Mathematische Forschungsdateninitiative.

\section*{Author contributions}

B.\ V.\ initiated the research. Cs.\ K.\ and B.\ V.\ derived the model, prepared the scripts and figures, and carried out the estimations. Cs.\ K., L.\ N.\ and B.\ V.\ analyzed the results and wrote the manuscript.

\section*{Data availability}

The titleholder data are freely available at https://grg.org/Adams/C.HTM

\section*{Additional information}

The authors declare no competing interests.

\section*{Tables}

\begin{table}[ht]
\centering
\[\begin{array}{|c|ccc|}
\hline
&E=0&E=30&E=60\\
\hline
C_E&6270&4.680\cdot10^{-9}&3.249\cdot10^{-11}\\
\kappa_E&0.004987&0.01876&0.02085\\
\hline
\end{array}\]
\caption{Birth rate parameters in \eqref{lambdat} for different values of the entry age $E$.\label{table:birthrate}}
\end{table}

\begin{table}[ht]
\centering
\[\begin{array}{|c|ccc|}
\hline
&E=0&E=30&E=60\\
\hline
\alpha_E&0.01277&0.01124&0.01110\\
K_E&0.00002951&0.0005950&0.01208\\
b&0.09&0.09&0.09\\
\gamma_E&0.08596&0.08061&0.08026\\
l(\alpha_E,K_E,0.09,0,\gamma_E)&-119.64&-117.68&-117.27\\
\hline
\end{array}\]
\caption{The optimal parameters obtained for $b=0.09$ and for various values of the entry age $E=0,30,60$.\label{table:optparam}}
\end{table}

\begin{table}[ht]
\[\begin{array}{|c|ccc|}
\hline
&E=0&E=30&E=60\\
\hline
\alpha_E&0.01561&0.01376&0.01352\\
K_E&3.728\cdot10^{-6}&0.0001495&0.005901\\
b&0.11&0.11&0.11\\
\gamma_E&0.1160&0.1117&0.1110\\
l&-119.48&-117.51&-117.09\\
\hline
\end{array}\quad
\begin{array}{|c|ccc|}
\hline
&E=0&E=30&E=60\\
\hline
\alpha_E&0.01845&0.01628&0.01596\\
K_E&4.530\cdot10^{-7}&3.613\cdot10^{-5}&0.002784\\
b&0.13&0.13&0.13\\
\gamma_E&0.1441&0.14065&0.1399\\
l&-119.37&-117.40&-116.98\\
\hline
\end{array}\]
\[\begin{array}{|c|ccc|}
\hline
&E=0&E=30&E=60\\
\hline
\alpha_E&0.02129&0.01880&0.01842\\
K_E&5.362\cdot10^{-8}&8.500\cdot10^{-6}&0.001281\\
b&0.15&0.15&0.15\\
\gamma_E&0.1708&0.1681&0.1674\\
l&-119.31&-117.34&-116.91\\
\hline
\end{array}\]
\caption{The optimal parameters for $b=0.11$, $b=0.13$ and $b=0.15$.\label{table:differentb}}
\end{table}

\end{document}